# Stellar Imager (SI): developing and testing a predictive dynamo model for the Sun by imaging other stars

## 10 November 2010

Kenneth G. Carpenter (GSFC), Carolus J. Schrijver (LMATC), Margarita Karovska (CfA),
Steve Kraemer (CUA), Richard Lyon (GSFC), David Mozurkewich (Seabrook Eng.),
Vladimir Airapetian (GMU), John C. Adams (GSFC), Ronald J. Allen (STScI),
Alex Brown (UCO/Boulder), Fred Bruhweiler (CUA), Alberto Conti (STScI),
Joergen Christensen-Dalsgaard (U. Aarhus), Steve Cranmer (CfA),
Manfred Cuntz (U. Texas/Arlington), William Danchi (GSFC), Andrea Dupree (CfA),
Martin Elvis (CfA), Nancy Evans (CfA), Mark Giampapa (NSO/NOAO),
Graham Harper (UCO/Boulder), Kathy Hartman (GSFC), Antoine Labeyrie (College de France),
Jesse Leitner (GSFC), Chuck Lillie (NGST), Jeffrey L. Linsky (UCO/Boulder),
Amy Lo (NGST), Ken Mighell (NOAO), David Miller (MIT), Charlie Noecker (BATC),
Joe Parrish (Aurora Flight Systems), Jim Phillips (CfA), Thomas Rimmele (NSO),
Steve Saar (CfA), Dimitar Sasselov (CfA/Harvard), H. Philip Stahl (MSFC),
Eric Stoneking (GSFC), Klaus Strassmeier (AI-Potsdam), Frederick Walter (SUNY),
Rogier Windhorst (ASU), Bruce Woodgate (GSFC), Robert Woodruff (LMSSC)

**For additional information, please contact:**

Dr. Kenneth G. Carpenter
Code 667, NASA-GSFC
Greenbelt, MD 20771
Phone: 301-286-3453, Email: **Kenneth.G.Carpenter@nasa.gov**

## *I. Summary of Science Mission Concept*

***The objective of the Stellar Imager (SI) mission is to accelerate the development and validation of a predictive dynamo model for the Sun and other magnetically active stars.*** In particular, we wish to understand the origins of variability in the Sun-Earth system and provide the foundations for more accurate long-term forecasting of solar/stellar magnetic activity and its impact on space weather, planetary climates, and life. To accomplish this, SI will:

- Characterize the patterns of surface magnetic activity for a large sample of Sun-like stars by obtaining time-resolved images with ~1000 resolution elements on their surfaces.
- Characterize the internal structure and differential rotation of these stars using asteroseismology with at least 30 resolution elements across the stellar diameter.
- Determine the dependence of dynamo action on mass, internal structure, flow patterns, and time by carrying out a population study of Sun-like stars. *This will enable testing of dynamo models over a few years of observations of many stars, instead of over many decades using only the Sun.*

Our understanding of the solar dynamo has been advanced significantly by combining solar observations with numerical experiments. Yet, none of the current models yields the Sun's average activity level; none of the self-contained numerical models has proven predictive capabilities; and the (quasi-)empirical models disagree strongly on forecasts for the next solar cycle. To make significant progress we really need to know what is important in making a dynamo go, i.e., the dependence of dynamo action on, e.g., mass, effective temperature, surface gravity, rotation rate, internal structure, flow patterns, and time. By choosing the correct stars for a population study we can vary one parameter at a time and see how that affects stellar activity.

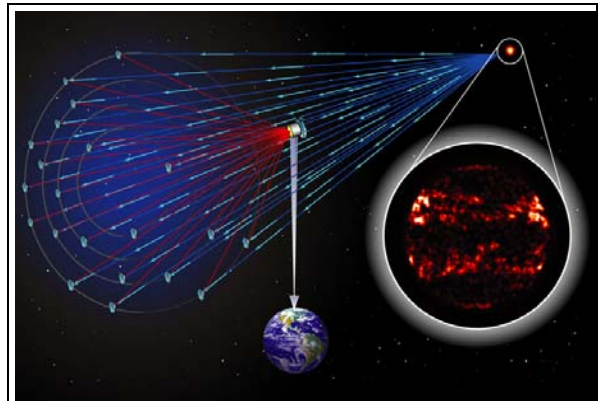

Fig. 1: SI - An array of 20-30 one-meter mirrors fly in precision formation to form a virtual parabola ~500m in diameter to enable sub-milliarcsecond spectral imaging of solar-type stars, e.g., as shown above in the light of CIV emission lines.

The Stellar Imager mission concept (Fig. 1) that will accomplish these goals is a space-based, UV/Optical Interferometer (UVOI). The required resolution dictates baselines ~500 meters, the imaging quality requires ~30 apertures, and the imaging rate requires apertures ~1.0 meter diameters. All three of these numbers can be decreased at the expense of a decreased number of targets and increased time to complete an image. The resulting instrument will have over 200x the spatial resolution of Hubble Space Telescope. It will enable 0.1 milli-arcsec (@2000Å) spectral imaging of stellar surfaces and, via spatially-resolved asteroseismology, measurements of the structure and differential rotation of stellar interiors.

The required technologies all have mid-level TRLs; the concepts are thoroughly tested, at least in the lab. Many have been deployed at ground-based astronomical observatories. ***However, to enable the SI mission in the late 2020's, significant technology development in the upcoming (2013+) decade is critical to qualify these technologies for space missions.*** The key technology needs include: 1) precision formation flying of many spacecraft and 2) closed-loop control of many-element, sparse optical arrays.

SI has been in the SEC/SSSC/Heliophysics Division Roadmaps since 2000 and is a Flagship "Landmark/Discovery Mission" in the 2005 Roadmap (Fig. 2). It was selected as a NASA Vision Mission (VM; "NASA Space Science Vision Missions" 2008, "Progress in Astronautics and Aeronautics", vol. 224, pp. 191-227; AIAA, ed. M. Allen). The full SI Vision VM Study report, related science and technology whitepapers, and additional information and references can be found at: http://hires.gsfc.nasa.gov/si/. Here we first discuss the primary science goals of SI in further detail, and then outline the mission architecture, technology needs, and estimated costs.

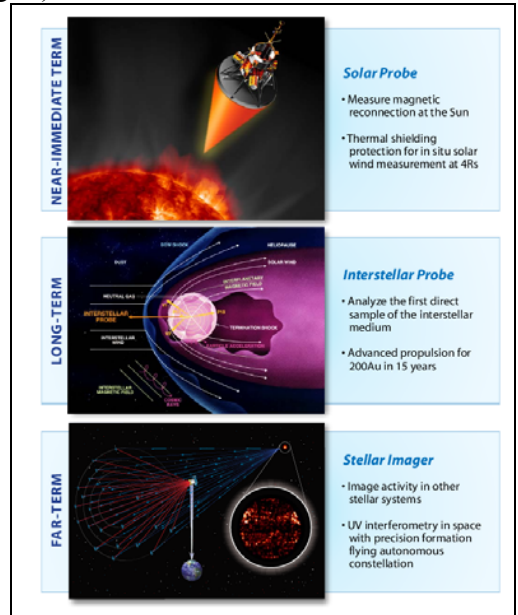

Fig. 2: Stellar Imager is identified as a "Landmark/Discovery Mission" in the 2005 SSSC (Heliophysics Division) Roadmap.

*A. Advancing Solar/Space Physics by Improving our Understanding of Dynamos and Magnetic Activity in the Sun & Stars*

A magnetic dynamo operates in all stars at least during some phases in their evolution. It regulates the formation of stars and their planetary systems, the habitability of planets, and the space weather around them. Dynamos also operate in objects including planets, accretion disks, and AGN. Although we know that flows in rotating systems are essential to such dynamos, there is no comprehensive dynamo theory from which we can derive the strength, the patterns, or the temporal behavior of stellar magnetic fields. Understanding the complex of non-linear couplings in a dynamo requires that we combine numerical studies and theory with observations of the evolving surface field of Sun and stars, of the average properties of magnetic fields in a large sample of stars, and of stellar internal flows. *These observations will enable us to constrain the source and sink properties of the surface magnetic field, by measuring the latitudinal and longitudinal flux-emergence patterns, the flux dispersal rate, and the differential and meridional flows. These properties are both important input parameters to flux-transport models and sensitive differentiators between dynamo models.*

For cool stars like the Sun, dynamo action persists from the very formation of the star throughout its existence as a fusion reactor. The Sun's activity is modulated significantly from cycle to cycle, sometimes persistently for several decades. Activity decreased, for example, for decades in the 17th Century when Earth experienced the Little Ice Age. Observations of other cool stars are crucial to our understanding of solar/stellar dynamo action. This has taught us, for example, that convection is part of all solar-like dynamos, and that rotation regulates their strength. But we do not have a theory that explains why stars are as active as they are, why some show cycles and some do not, what causes the Sun's cycles to differ from one to the next (e.g., the extra long duration of Cycle 23), and how a cyclic dynamo can restart after a Maunder-like (e.g. Little Ice Age) minimum. Hence, we cannot usefully forecast long-term space weather or reliably model the effects of stellar magnetism on the evolution of stars, planetary systems, planetary atmospheres, and thus the habitability of a planetary system.

Numerical modeling has taught us valuable lessons about dynamos, including the fact that they are highly non-linear and couple processes that occur over geometric extents as large as the full convection zones of stars down to the smallest convective scales. Our computer models are therefore of necessity simplified, and hence require observational guidance from stars with a variety of properties. *Key to successfully developing a predictive dynamo theory is the realization that we need a population study of stars at high spatial resolution: we need to study the evolution of dynamo-driven activity in both latitude and longitude in a sample of stars like the Sun, and compare it to observations of young stars, old stars, binary stars, etc.* The potential for a breakthrough in our understanding lies in spatially-resolved imaging of the dynamo-driven emission patterns on this wide-ranging stellar sample. These patterns, and how they depend on stellar properties (such as convection, (differential) rotation and helicity, meridional circulation, and evolutionary stage/age), are crucial for dynamo theorists to explore the sensitive dependences on many poorly known parameters, to investigate bifurcations in a non-linear 3D dynamo, and to ultimately validate a model. *Direct UV/optical interferometric imaging* (0.1 mas observations in UV emission lines of magnetically active regions on stellar surfaces) *is the only way to obtain the required information on the dynamo patterns for stars of Sun-like activity.*

To address these science goals, the Stellar Imager is designed to provide UV/Optical sub-milliarcsec images and disk-resolved asteroseismology for a sample of stars similar to the Sun, as well for other cool stars with very different characteristics. SI will have access to an exciting array of distinct stars and stellar systems (see Fig. 3). An array diameter of 500 m is needed to resolve a medium-sized solar-type active region when observing a Sun-like star at ~4 pc. A km-sized array provides the necessary resolution out to ~8 pc. SI will, for the first time, enable imaging of magnetic activity of a variety of Sun-like stars (there are ~3 dozen F, G, K main-sequence (MS) stars within 8 pc) and many cooler M-type MS stars, including many nearby and more distant stars with shallow convective envelopes, fully-convective stars, close binary systems with dual active components magnetically coupled at a few stellar radii, compact RS-CVn-type binaries, mass-transferring Algol-type systems, symbiotic systems, and giant & supergiant stars. Imaging magnetically active stars and their surroundings will also provide us with an indirect view of the Sun through time, from its formation in a molecular cloud, through its phase of decaying activity, during and beyond the red-giant phase during which the Sun will swell to about the size of the Earth's orbit, and then toward the final stages of its evolution (see Fig. 4).

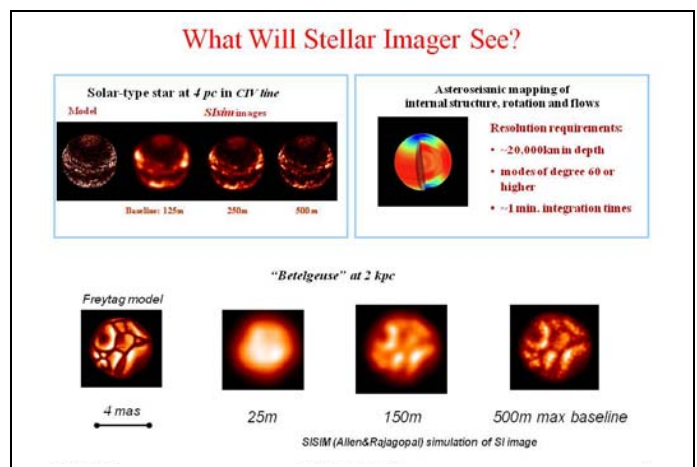

Fig. 3: SI will provide UV images of stars at resolutions only previously dreamed of, as seen in these simulations (using 30 mirrors in a non-redundant pattern with indicated max. baselines).

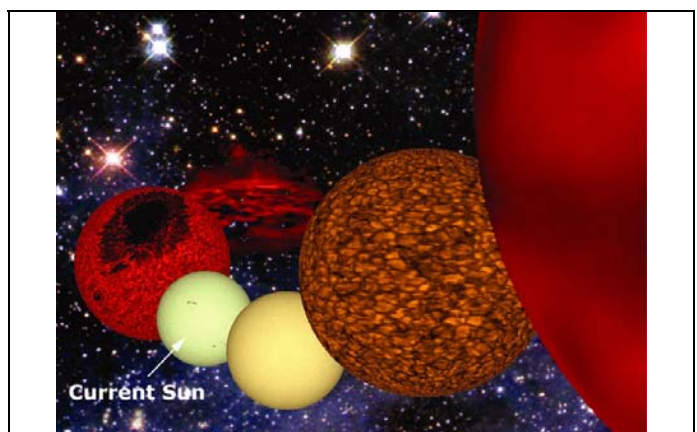

Fig. 4: Evolution of the Sun in time - SI will provide us a view by resolving stars representing each solar era.

*B. Mission Architecture*

NASA commissioned a "Vision Mission (VM) Study" of SI in 2004-2005, which developed a design that meets the performance requirements required to achieve the science goals described above. The baseline full-mission concept for SI was developed in collaboration with the GSFC Mission Design Lab (MDL) and Instrument Design Lab (IDL). The MDL worked on the overall design of a space-based Fizeau interferometer, located in a Lissajous orbit around the Sun-Earth L2 point. A variety of disciplines considered the implications of this general design, including power, guidance & navigation, flight dynamics, operations, communications, quality assurance, system engineering, etc. The IDL concentrated its efforts on the design of the beam-combining hub in the context of the selected overall architecture, again from a multiple-discipline viewpoint, and including accommodation of the MDL results. In addition to assisting in the development of the architecture, the Design Labs explored the technical feasibility of the mission and identified the technology developments needed to enable the mission in the late 2020's. In order to address its science goals, SI must have the following capabilities:

- **wavelength coverage: 1200 – 6600 Å**
- **access to UV emission lines** from Ly$\alpha$ 1216 Å to Mg II 2800 Å
    - Important diagnostics of most abundant elements
    - much higher contrast between magnetic structures and background
    - shorter baselines (UV saves 2-4x vs. optical, active regions 5x larger)
    - ~10-Å UV pass bands, e.g. C IV (100,000 K); Mg II h&k (10,000 K)
- **broadband, near-UV or optical** (3,000-10,000 K) for high temporal resolution spatially-resolved asteroseismology to resolve internal stellar structure
- angular resolution of 50 $\mu$as at 1200 Å (120 $\mu$as @2800 Å) to provide ~1000 pixels of resolution over the surface of nearby (4pc) dwarf stars, and more distant giant and supergiant stars.
- **spectral resolution** of R>100 (min) up to R=10000 (goal)
- **long-term (~ 10 year) mission**, to enable study of stellar activity cycles:
    - individual telescopes/hub(s) can be refurbished or replaced

The VM study defined a detailed flow down of requirements from science goals to data and measurements requirements to engineering implications to the key technologies needed to implement the mission. The baseline mission architecture shown in Fig. 5 was derived from these requirements. The selected design is a space-based (at Sun-Earth L2), UV-Optical Fizeau Interferometer with 30 one-meter primary mirrors, mounted on formation-flying "mirrorsats" distributed over a parabolic virtual surface whose diameter can be varied from 100m up to as much as 1000m, depending on the angular size of the target to be observed. Table 1 summarizes the mission and performance parameters of the baseline SI design.

*C. Technology Drivers*

The two major technology challenges (the "tallest poles") to building SI are:

- **precision formation-flying (PFF) of ~20-30 spacecraft** (including precision metrology over baselines of 100's of meters)

- **wavefront sensing and real-time autonomous analysis and closed-loop optical control of a many-element sparse array**

These technology challenges have been addressed, along with the less difficult ones, prior to and during the SI Vision Mission (VM) study and in GSFC Integrated Design Center (IDC) studies.

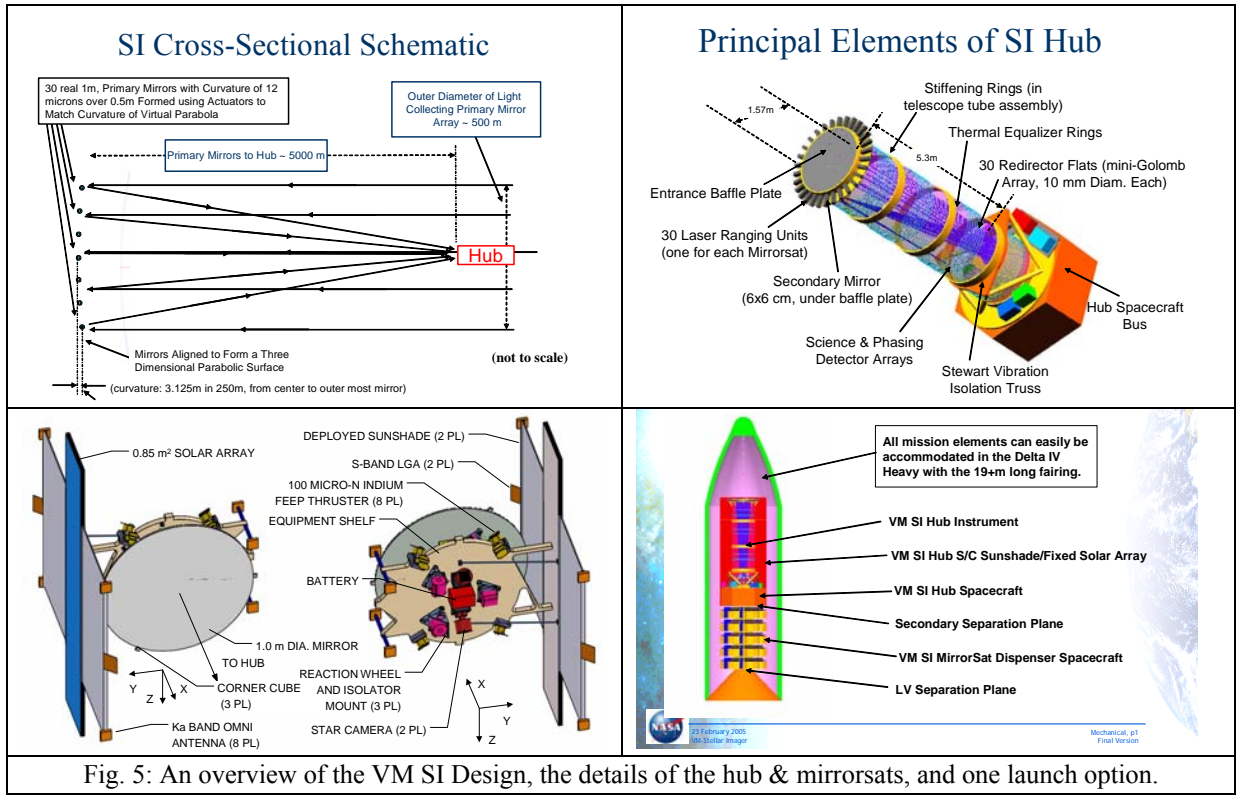

Fig. 5: An overview of the VM SI Design, the details of the hub & mirrorsats, and one launch option.

**Table 1: Mission and Performance Parameters from the Vision Mission Study**

| Parameter | Value | Notes |
|---|---|---|
| Maximum Baseline (B) | 100 – 1000 m (500 m typical) | Outer array diameter |
| Effective Focal Length | 1 – 10 km (5 km typical) | Scales linearly with B |
| Diameter of Mirrors | 1 - 2 m (1 m currently) | Up to 30 mirrors total |
| λ-Coverage | UV: 1200–3200 Å ; Opt: 3200–6600 Å | λ-Sensing in optical |
| Spectral Resolution | UV: 10 Å (emission lines) UV/Opt: 100 Å (continuum) | |
| Operational Orbit | Sun-Earth L2 Lissajous, 180 d | 200,000x800,000 km |
| Operational Lifetime | 5 yrs (req.) – 10 yrs (goal) | |
| Accessible Sky | Sun angle: 70º ≤ β ≤ 110º | Entire sky in 180 d |
| Hub Dry Mass | 1455 kg | For each of 2 |
| Mirrorsat Dry Mass | 65 kg (BATC design) | For each of 30 |
| Ref. Platform Mass | 200 kg | |
| Total Propellant Mass | 750 kg | For operational phase |
| Total Mass to Orbit | 4355 kg | |
| Class of Launcher | Delta IV Heavy – 19 m fairing | |
| Angular Resolution | 50 µas – 208 µas (@1200–5000Å) | Scales linearly ~ λ/B |
| Typical time to image star | < 5hr for solar type; < 1 day for supergt, | |
| Imaging time resolution | 10 – 30 min (10 min typical) | Surface imaging |
| Seismology time res. | 1 min cadence | Internal structure |
| # res. pixels on star | ~1000 total over disk | Solar type at 4 pc |
| Minimum FOV | > 4 mas in single exposure | Larger via mosaic |
| Minimum flux detectable at 1550 Å | $5.0 \times 10^{-14}$ ergs/cm$^2$/s integrated over C IV lines | 10 Å bandpass |
| Precision Formation Fly. | s/c control to mm-cm range | |
| Optical Surfaces Control | Actuated mirrors to µm-nm range | |
| Phase Corrections | to λ/10 Optical Path Difference | |
| Aspect Control/Correct. | 3 µas for up to 1000 sec | Line of sight mainten. |

Credible and feasible approaches to the successful development of all these technologies were derived during the course of those studies and are documented in the SI VM Final Report, but these must be funded and implemented during the 2013 decade to enable the technologies to be ready for flight in the following decade. *All of these technologies exist – it is simply a matter of extending such work to improve performance and/or operational scale.* Various ground-based experiments, e.g., the MIT SPHERES, GSFC FFTB, and JPL FCT formation flying testbeds and GSFC Fizeau Interferometer (large optical array phasing/synthetic imaging) Testbed, can be used to mature these technologies in preparation for validation on a small, space-based experiment (for PFF) and on ground-based optical interferometers such as MROI (for optical array phasing).

## *II. Cost Estimates*

We estimate the total cost, over the full decade, of the ground-based technology work described above, at ~$37M. Costs are estimated based on experience-to-date with the efforts noted above. Flight validation of PFF technologies is also required by other potential missions and would build on the results of the Orbital Express (DARPA) and PRISMA (Swedish Sp. Corp.) missions and the upcoming PROBA-3 (ESA) and F-6 (DARPA) missions. If an additional mission becomes necessary, it is expected to be funded by cross-enterprise Strategic Technology Development initiatives, with additional cost-sharing possible from DoD and/or ESA sources. *An endorsement by the Decadal Survey of the SI science and mission as worthy of flight in the following (2023) decade would enable us to pursue funding in the 2013 decade for the technology development, mission design, and international coordination*.

Over the last 8 years we have performed a number of SI design studies in the GSFC IDC, including studies of the overall mission, the beam-combining hub, and the science payload. These studies involved experts from a full range of technical disciplines, including power, guidance & navigation, flight dynamics, operations, communications, quality assurance, system engineering, etc., and cost estimation. These studies explored the technical feasibility of the mission, identified the technology developments needed to enable the mission in the 2024+ timeframe, and provided cost estimates as described below. We have used both Parametric and Analogous methods to estimate the cost of the SI mission. Parametric Cost Modeling at GSFC includes the use of PRICE-H (Hardware) and/or SEER-H (Hardware) for mission hardware cost estimating. The Analogous cost methodology is based on historical data and involves comparison and extrapolation to like items or efforts in previous missions and design efforts.

We first considered the costs of the mirrorsats and beam-combining hub, using both Price-H and the JSC mass-based Advanced Missions Cost estimator (http://cost.jsc.nasa.gov/AMCM.html). Costs based on Price-H from the GSFC IDC sessions were used and inflated to 2009$. The mirrorsats were studied during 2001, the hub spacecraft in 2004, and the payload in 2005. Second, the spacecraft/vehicle level mass-based cost estimator from the JSC web site (http://cost.jsc.nasa.gov/SVLCM.html) was used assuming the parameters shown in Table 1. We also compare in Table 2 the cost estimations for the mirrorsats and hub/payload. Inflating the JSC 2004$ estimates to 2009$ gives $1.5B which is approximately 36% higher than the $1.1B cost for mirrorsats and hub/payload based on Price-H estimates. The Price-H estimates were used in the total mission

**Table 2. Cost Comparisons for Mirrorsats and Hub/Payload ('09$)**

| Component | Price-H/IDC | JSC Estimator (avg. complexity) | Assumptions |
|---|---|---|---|
| 30 Mirrorsats | $656M | $759M | 65 kg each |
| One Payload Hub | $466M | $767M | 1455 kg |
| Component Total | $1.1B | $1.5B | |

cost derived below, since they were generated using estimates of specific hardware component costs and we believe them to be more reliable.

Project management, mission systems engineering, mission assurance, and system integration and testing are estimated as percentages of hardware costs based on cost averages from the extensive GSFC mission flight experience. Science costs in Phases C/D are based on $500K per instrument per year for a flagship mission class, while Phase E is scaled from Hubble Space Telescope (HST), accounting for fewer discrete targets and programs/year. Four instruments are included: UV and visible cameras, a wavefront sensing and control system, and a "light bucket" spectrograph, which are costed based on HST experience. In Phase E, $ 15 M/year (of line 4.0) goes directly to the astronomical community in the form of grants. Reserves are computed as 30% of Phase B-E cost before adding the launch vehicle cost. EPO costs are based on appropriate levels for a Flagship mission. We selected a Delta-IV Heavy as an example launch vehicle, and very conservatively estimated its cost from information on the Astrophysics Strategic Mission Concept Studies workshop webpage, i.e., 4.2 times the cost of a small launch vehicle ($160M in 2015$), and deflated to 2009$. The SI hardware can be accommodated in other launch vehicles with similar characteristics.

The process above yields a total mission cost in 2009$ of $2.9B. *We note that there is significant international interest (Carpenter et al., 2009 Ap & Sp. Sci., 320, 217) in the SI mission as well as from the Astrophysics community, which is interested both in the stellar astrophysics and other topics addressable by a high resolution UV/Optical observatory (e.g., central engines of AGNs, mass accretion processes, dynamical structures in supernovae and planetary nebulae) and we would thus expect this cost to be shared by international (esp. European) partners and by the Astrophysics Division.* The breakout and methodologies used to estimate this mission cost are shown in Table 3. We have left our formal estimates in FY09$, as produced by our last major costing exercise, since the current extremely low level of inflation would not materially change the result relative to other uncertainties in the costing.

**Table 3: SI Mission Cost Estimate ($ M, 2009) and Methodology - updated 3/23/09**

| Cost Element | Phase A | Phase B | Phase C/D | Dev Total | Phase E | Mission Total | Cost Methodology |
|---|---|---|---|---|---|---|---|
| **Project Elements:** | | | | | | | |
| 1.0 Project Management | 2 | 9 | 91 | 102 | 9 | **111** | 9% hardware costs B-D;9% of Mission Ops |
| 2.0 Mission Sys Engr | 2 | 8 | 81 | 92 | 8 | **100** | 90% of Project Mgmt |
| 3.0 Mission Assurance | 2 | 3 | 33 | 38 | 3 | **41** | 40% of Systems Engineering |
| 4.0 Science | 2 | 3 | 6 | 11 | 129 | **140** | Based on Flagship mission data; analogous ops costs |
| 5.0 Payload & SC | 5 | 52 | 409 | 466 | 0 | **466** | MDL and IDL estimates, Price-H |
| 6.0 Flight Systems (incl. 30 mirrorsats) | 6 | 52 | 598 | 656 | 0 | **656** | MDL and IDL estimates, Price-H |
| 7.0 Mission Ops | 1 | 4 | 15 | 21 | 103 | **124** | MDL estimates; analogous cost, and cost calculators |
| 9.0 Ground System | 2 | 2 | 2 | 6 | 10 | **16** | MDL estimates; and analogous cost |
| 10.0 System I&T | 1 | 10 | 101 | 112 | 0 | **112** | 10% of hardware costs |
| **Sub total** | **24** | **143** | **1336** | **1504** | **263** | **1766** | |
| Reserves | 0 | 43 | 401 | 444 | 64 | **508** | |
| **Sub total w/reserves** | **24** | **186** | **1737** | **1948** | **327** | **2273** | |
| **Elements w/o cont:** | | | | | | | |
| 8.0 Launch Vehicle | 0 | 0 | 571 | 571 | 0 | **571** | Delta IV H cost from ASMCS webpg. |
| 11.0 E/PO | 0 | 3 | 12 | 15 | 15 | **31** | Phase B-E, appropriate for Flagship |
| **Mission Total:** | **24** | **190** | **2320** | **2534** | **342** | **2875** | |